# The on-board data handling concept for the LOFT Large Area Detector

S. Suchy[a], P. Uter[a], C. Tenzer[a], A. Santangelo[a], A. Argan[b], M. Feroci[c], T.E. Kennedy[d], P.J. Smith[d], D. Walton[d], S. Zane[d], J. Portell[e,f], E. García–Berro[e,g]
On Behalf of the LOFT Consortium

[a] Institute for Astronomy and Astrophysics, Eberhard Karls University Tübingen, Sand 1, 72076 Tuebingen, Germany;
[b] INAF HQ, Viale del Parco Mellini 84, 00136, Roma, Italy;
[c] INAF/IASF Roma, via Fosso del Cavaliere 100, 00133, Roma, Italy;
[d] Mullard Space Sciences Laboratory, University College London, Holmbury St. Mary, Dorking, Surrey RH5 6NT, UK
[e] Institute for Space Studies of Catalonia (IEEC), C/ Gran Capit 2-4, 08034 Barcelona, Spain
[f] Dept. Astronomy and Meteorology, University of Barcelona, C/ Mart i Franqus 1, 08028 Barcelona, Spain
[g] Departamamente de Física Aplicada, Universitat Politècnica de Catalunya, c/Esteve Terrades 5, 08860 Castelldefels, Spain

## ABSTRACT

The Large Observatory for X-ray Timing (LOFT) is one of the four candidate ESA M3 missions considered for launch in the timeframe of 2022. It is specifically designed to perform fast X-ray timing and probe the status of the matter near black holes and neutron stars. The LOFT scientific payload consists of a Large Area Detector and a Wide Field Monitor.

The LAD is a $10\,\text{m}^2$-class pointed instrument with high spectral (200 eV @ 6 keV) and timing ($< 10\,\mu$s) resolution over the 2-80 keV range. It is designed to observe persistent and transient X-ray sources with a very large dynamic range from a few mCrab up to an intensity of 15 Crab.

An unprecedented large throughput (∼280.000 cts/s from the Crab) is achieved with a segmented detector, making pile-up and dead-time, often worrying or limiting focused experiments, secondary issues.

We present the on-board data handling concept that follows the highly segmented and hierarchical structure of the instrument from the front-end electronics to the on-board software. The system features customizable observation modes ranging from event-by-event data for sources below 0.5 Crab to individually adjustable time resolved spectra for the brighter sources. On-board lossless data compression will be applied before transmitting the data to ground.

**Keywords:** LOFT, LAD, electronics, data handling

## 1. INTRODUCTION

High-time-resolution X-ray observations of compact objects provide direct access to strong-field gravity, black hole masses and spins, and the equation of state of ultradense matter. A large instrument in combination with good spectral resolution is required to exploit the relevant diagnostics and answer the two fundamental questions of ESAs Cosmic Vision Theme "Matter under extreme conditions": What is the fundamental equation of state (EOS) of a compact object? Does matter orbiting close to the event horizon follow the predictions of general relativity?

suchy@astro.uni-tuebingen.de

## 1.1 Scientific Goals

Understanding the properties of ultradense matter and determining its EOS is one of the most challenging problems in contemporary physics. A very soft EOS gives a maximum neutron star mass in the 1.4-1.5 solar mass ($M_\odot$) range, whereas a 'stiff' EOS can give a neutron star mass of up to 2.4 -2.5 $M_\odot$ before a collapse to a black hole becomes unavoidable. Apart from maximum mass, the relation between the neutron star mass and radius ($M_{NS}$-$R_{NS}$) is a powerful probe of the EOS. In $\sim$25 neutron stars, spins are now observed in burst oscillations and/or coherent pulsations at frequencies of up to 620 Hz, proving that millisecond spins and dynamically relevant magnetic fields are common among neutron stars in low-mass X-ray binaries. LOFT will measure the masses and radii of accreting millisecond pulsars to an accuracy of 4% and 2-3%, respectively, by modelling their pulse profiles. Since the maximum rotation that a neutron star can sustain depends on its mass and structure, fastest spin periods also constrain the neutron star EOS. A different approach has recently emerged from the discovery of global seismic oscillations (GSOs) in the tens of Hz to kHz range from magnetars during the rare and extremely luminous giant flares emitted by these sources. The lower frequency GSOs likely arise from torsional shear oscillations of the crust and their frequency, in combination with the magnetic field inferred from the magnetar spin-down, tightly constrains the EOS. LOFT will detect and study GSOs for the first time in intermediate flares, which are tens of times more frequent than giant ones, down to amplitudes of an order of magnitude lower than observed up to now.

About 40 compact objects accreting matter in binaries are now known to display variability arising in, and occurring at the millisecond dynamical timescale of their inner accretion flows. Black holes and neutron stars show Quasi-Periodic Oscillations (QPOs) of up to 450 and 1250 Hz, respectively. These QPOs require an explanation that involves the fundamental frequencies of the motion of matter in the inner, strong-field gravity-dominated disk regions. Very high signal-to-noise LOFT/LAD measurements of the QPOs will unambiguously discriminate between such interpretations and in the process probe yet untested general relativistic effects, such as strong-field periastron precession, the presence of an innermost stable orbit and frame dragging. LOFT will allow to measure dynamical timescale phenomena within their coherence time for the first time, where so far only statistical averages of signals were accessible. It will also allow direct measurements of the black hole mass and spin through timing measurements, to compare with other estimates such as mass from optical studies or spin from the thermal X-ray continuum or the Fe K-line profile. With the spectral and timing capabilities of LOFT it will be possible to measure the compact object's mass and spin, the disk inclination and to study the massive black holes in active galactic nuclei (AGN) with unprecedented accuracy by analyzing the pulse profiles and variability of their Fe K-lines.

In addition, LOFT will be a powerful tool to study the X-ray variability and spectra of a wide range of known objects, from accreting pulsars and bursters, to magnetar candidates (Anomalous X-ray Pulsars and Soft Gamma Repeaters), cataclysmic variables, bright AGNs, X-ray transients and the early afterglows of Gamma Ray Bursts. Through these studies it will be possible to address a variety of problems in the physics of these objects. Coordinated optical/NIR and radio campaigns on specific themes, as well as spin measurements which can aid the Advanced Virgo/LIGO searches for gravitational wave signals from fast rotating neutron stars will add great value to the LOFT program.

## 1.2 LOFT spacecraft

The Large Observatory For X-ray Timing (LOFT)[1] achieves an effective area of $> 10\,\mathrm{m}^2$ in the 2-30 keV range, yet still fits a conventional platform of a small/medium-class launcher (i.e. the ESA Soyuz launcher). The two main instruments are the Large Area Detector (LAD)[2] and the Wide Field Monitor (WFM).[3] The nominal LOFT mission duration is 4 years. The satellite will be launched into a low equatorial orbit (LEO) with an altitude of $\sim 600\,\mathrm{km}$ and an inclination of $< 5°$. Contact to the spacecraft will be performed each orbit by two dedicated ground station at low latitude. Figure 1 shows an overview over the LOFT spacecraft with its individual components labeled.

The key to the large increase in effective area in the LAD ($> 10\times$ area over previous missions) by LOFT resides in the synergy between technologies imported from other fields of scientific research, both ground and space-based. The crucial characteristic of the LAD is a mass per unit surface in the range of $\sim 30\,\mathrm{kg\,m^{-2}}$, enabling an $\sim 18\,\mathrm{m}^2$ geometric area payload at reasonable weight. This is achieved by the use of sensitive but light large-area

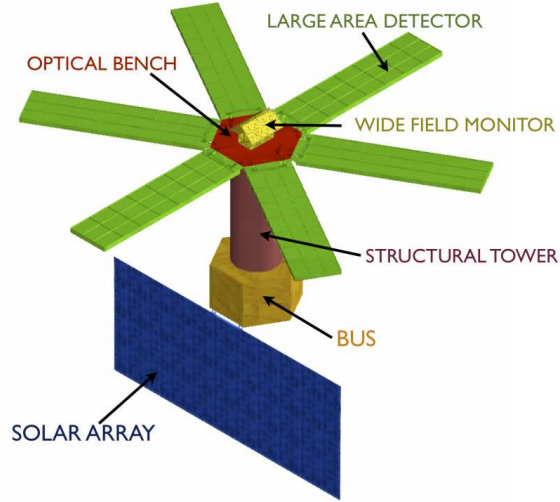

Figure 1. Schematic view of the LOFT spacecraft with its individual elements labeled.

Silicon Drift Detectors (SDDs), designed on the heritage of the ALICE experiment at CERN/LHC[4,5] and by light weight collimators produced from lead-glass microcapillary plates (based on the well known microchannel plates). The large detector panels of the LAD are deployed through a mechanism used in the Synthetic Aperture Radar missions (e.g., the SMOS ESA mission, in orbit since 2009), where similar large panels were deployed and pointed with high accuracy.

### 1.3 Large Area Detectors

The Large Area Detector (LAD) is designed as a classic collimator experiment with a field of view (FOV) of $\sim 1°$. The key properties of the SDD[6] are their capability to read-out a large photon collecting area with a small set of low- capacitance (thus low-noise) anodes and their very small weight ($< 1\,\mathrm{kg\,m^{-2}}$). The working principle is shown in Figure 2, where an incident photon generates a cloud of electrons in the detector. These electrons are then driven by a constant electric field to the read-out anodes. The diffusion in Si causes the electron cloud to expand by a factor depending on the square root of the drift time and therefore the drift distance. The charge distribution over the collecting anodes thus depends on the absorption point in the detector.

The LAD detector design is an optimization of the ALICE detector: $450\mu m$ thick wafers will be used to produce 76 cm$^2$ monolithic SDDs (Figure 3) with an anode pitch of 970 $\mu m$. The single silicon tile is electrically divided in two halves, with 2 series of 112 read-out anodes at two edges and the highest voltage along its symmetry axis. The maximal drift length is 35 mm. A drift field of 370 V/cm (1300 V maximum voltage), gives a drift velocity of $\sim$5 mm/$\mu$s and a maximum drift time of $\sim$7 $\mu$s. Depending on the relative size and position of the Gaussian-shaped charge cloud when reaching the anode pattern, the event charge may be collected by 1 or 2 anodes. Based on this, an event is defined as a single event ($\sim 45\%$ of all events), when the full charge is collected by a single anode, and a double event ($\sim 55\%$) when the charge is shared on two neighboring anodes. Due to the smaller noise, the single events display higher spectroscopic quality and can be selected for observations requiring higher spectral performance.

The LAD instrument is organized in a very hierarchical way. In the baseline configuration, the whole LAD instrument is divided into 6 individual panels looking at the same point in the sky. A LAD panel consists of 21 individual modules, each mounted in an aluminum box. The panel provides the mechanical frame and alignment interfaces, as well as the routing of the data signals and power supply for the individual modules. A panel backend electronics (PBEE) communicates with the individual detector modules, and reorganizes the measured data before it is transfered to the data handling unit (DHU) on the satellite bus. Each detector module consists of 16 individual SDDs, mounted in a 4 × 4 grid in a common enclosure. For each module, a module backend electronic (MBEE) is processing the data from each individual detector and forwards the processed event to the PBEE. An SDD is read out by 14 ASICs, where 7 ASICs are read out at one time. The LAD is highly modular and redundant, so that the loss of individual SDDs or detector modules does not significantly impact the scientific goals.

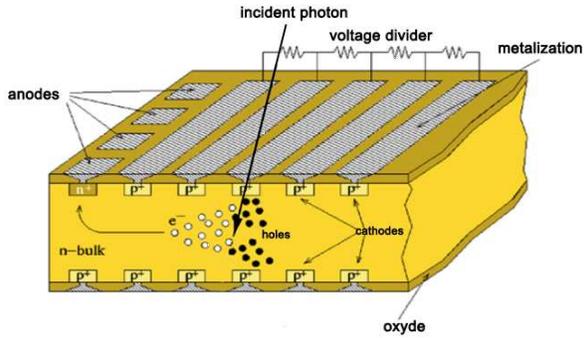
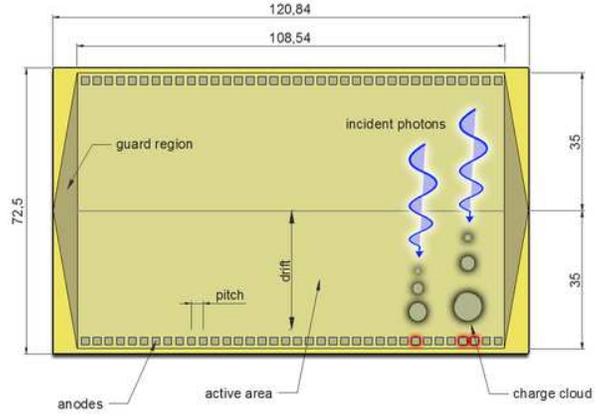

Figure 2. Working principle of a SDD, where the produced electrons are driven to the anodes for read-out.

Figure 3. Silicone drift detector design. A single SDD is divided symmetrical in the middle, and the charge cloud is moved to the respective anodes.

### 1.4 Data Flow

If an incident charge is measured above a pre-determined analogue trigger threshold in the ASIC a trigger signal is send to the MBEE, where the timestamp is frozen. The triggered ASIC transmits a trigger map, indicating which anodes measured a charge above their trigger thresholds. The MBEE validates the trigger map, and confirms that only a single anode or only two adjacent anodes actually triggered. After validation, the MBEE sends a command to the ASIC to either convert the stored charge to digital values or to discard the whole event and reset the anodes. After the event is either transferred to the MBEE or discarded, the SDD is ready to detect another event.

The MBEE processes the data by subtracting the known noise components and by reformatting the measured charge to an energy value (see Section 2). After the processing in the MBEE, the data from all 21 modules is transmitted to the PBEE, where it is sorted and reformatted. As part of the restructuring, the PBEE creates a low resolution energy spectrum in pre-determined time intervals, which can be used for either a quick-look result of the observed source, but also as a long time comparison of individual sources. After reconstruction, the data is finally transfered to the DHU on the satellite bus, where a lossless data compression is performed to reduce the amount of data to be transmitted to the ground station. The schematic view of the data flow is shown in Figure 4.

### 1.5 Telemetry requirements

The LAD scientific telemetry budget is estimated by assuming default event-by-event data transmission of 24-bit per event. Assuming an average intensity of 500 mCrab in the field of view the LAD would create an expected count rate of $\sim 120\,000$ cts/s for all SDDs. Taking into account the typical net source exposure in one low earth orbit ($\sim 4000$ s) a total of $\sim 11.5$ Gbit are created over one satellite orbit, corresponding to 1.9 Mbps orbit-average. Using a lossless data compression algorithm in the DHU, this amount can be compressed by a factor of $\sim 2$ using an algorithm adequately tailored to the features of the instrument, which may be based on an entropy coder such as Rice or PEC,[7] to reduce the data to $\sim 960$ kbps. A 64 GB mass memory on the DHU will allow the temporary storage of excess telemetry, also giving some flexibility if a connection between ground station and satellite is not successful. For sources with an LAD count above 120 000 cts/s, a flexible binned data mode will be implemented to reduce the overall amount of data.

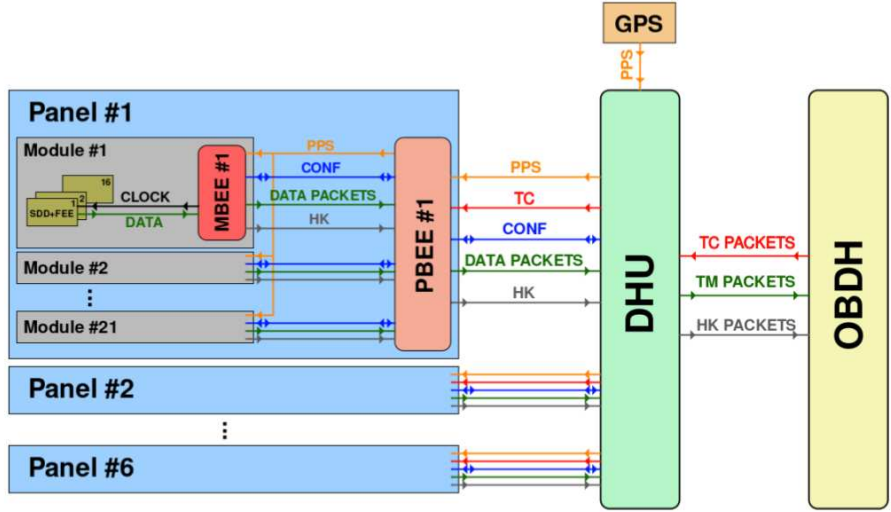

Figure 4. Block diagram showing the organizational structure in the LAD instrument.

The observing plan will be optimized by alternating bright and weak sources to allow for a gradual download of the excess telemetry, a strategy already successfully adopted by RossiXTE. Assuming two ground stations, the amount of data that can be transmitted to the ground per orbit is between 6.68 Gbit (S-band) and 10.58 Gbit (X-band). A higher transmission rate would allow to download stored data in orbits where the data rate was much higher (due to high intensity of the source) or where a data link was not achieved for different reasons (e.g. bad weather). Adopting new technical solutions to improve the down-link would provide the following advantages for the LAD operations and science return:

- full event info transmission also for sources of intensity higher than the current 500 mCrab baseline limit (see Section 3).

- unlimited and unconstrained observation of sources with intensity higher than 500 mCrab (lower or cancel any constraints on the sources observation scheduling based on their intensity).

- reduced complexity on: operations, data handling, mass memory, on-board software and ground software.

The assumptions for the observational limit of bright sources are based on the worst case scenario of an S-band transmission with only one available ground station.

## 2. MODULE BACKEND ELECTRONICS

The following section describes in more detail the individual tasks of the MBEE and the individual steps in the data pipeline.

The main tasks of the MBEE are:

- time tagging and event validation

- Offset and Common Noise correction

- Gain correction and energy addition

- Formation of event packets

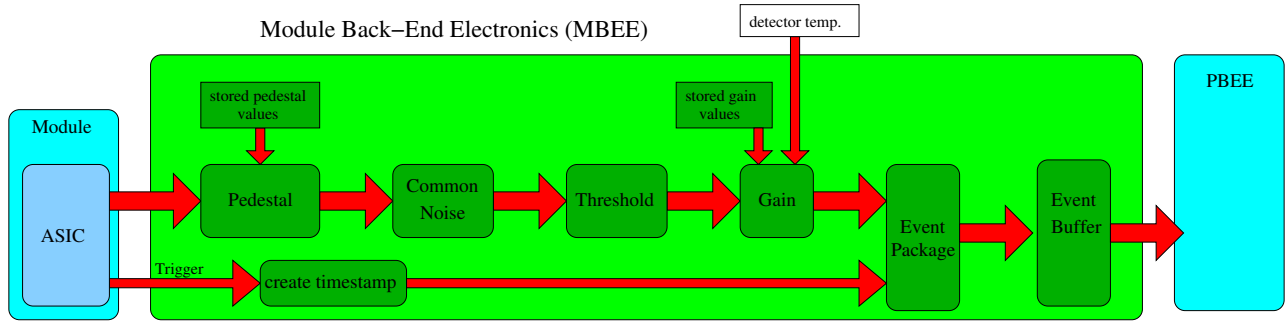

Figure 5. Schematic view of the LOFT data handling pipeline.

In addition, the MBEE measures and creates the housekeeping data which are monitored by the PBEE.

As mentioned above, 16 SDDs are combined to one detector module operated by a single MBEE. As the SDDs are divided electronically into two parts, the data is processed in 32 parallel data pipelines, which handle each detector half independently. Figure 5 shows a schematic description of such a data pipeline, where the individual steps of the MBEE are described in more detail below.

## 2.1 Event validation and time tagging

After an event occurs in the SDD, the charge cloud drifts towards the anodes and is measured. If the detected charge exceeds a pre-determined trigger level, the ASIC that connects to the specific anode sends immediately a trigger signal to the MBEE. After the shaping time, when the whole charge is accumulated in a buffer, the ASIC sends a created trigger map to the MBEE. The trigger map is a binary array of 16 bits indicating which anodes accumulated a charge above their trigger threshold. As 7 ASICs are measuring the anode charge on one side of the SDD, it can occur that an event is divided over adjacent ASICs. In such a case, the trigger map of both ASICs will be read out and validated. Only trigger maps where a single anode or two adjacent anodes have triggered are used for further analysis. If valid, The ASIC receives a command to convert the analog signal to an 11 bit ADC value, representing the accumulated charge, which is then transmitted to the MBEE. In cases where the trigger map is invalid, e.g. a high energy particle events that triggers multiple anodes, the analog values are deleted and the detector is cleared of charge to detect the next event. Avoiding the digital conversion of an invalid event reduces the dead time of the SDD significantly. In the time where an event is validated and converted, no other events can be measured in the detector half. As the conversion time for each event is the same, only two dead times occur, with and without event. The number of valid and invalid events are counted and are included in the housekeeping data.

## 2.2 Pedestal noise

The first step within each data processing pipeline is the pedestal subtraction of each anode. A set of pedestal values are stored for the anodes within the MBEE. These values have been determined during the detector calibration on the ground and can be updated after calibration observation in orbit or before each observation. The pedestal values are subtracted for each anode so that the zero value after the A/D conversion corresponds to allow for a linear energy gain correction later.

## 2.3 Common Mode Noise

Common mode noise (CN) in the anodes is primarily produced due to electronic noise in the anodes connected to the same ASIC. It is caused by an induced charge on all anodes and leads to an undesired baseline shift. Two main effects produce the CN component: the CN produced by the detector and the CN introduced by the electronic noise from the ASIC. This component will be calculated independently for each individual event, where the median value of all non-triggered anodes is assumed to be a good representation of the CN value in an event. Using the median instead of the average value allows to be less sensitive to individual noisy channels. The so determined CN value is subtracted from all anodes.

## 2.4 Gain Correction

After the CN correction, the energy of the event can be reconstructed. Each anode signal is corrected by an individual, linear gain factor which is stored in a lookup table in the MBEE. The initial values are determined during the detector calibration on the ground but can be updated in orbit. The temperature dependent variation in the energy gain is assumed to be $\sim 0.1\%$ per K temperature difference. As the gain is determined on the ground at a specific temperature and the temperature also will vary throughout the orbit, an onboard gain correction is unavoidable. Temperature sensors in each module will measure the actual detector temperature, which is then used for the gain correction. The linear correlation $E = E_{T_0} * (1 - C * (T - T_0))$ describes this variation, where $T_0$ is the temperature of the initial calibration, $T$ is the measured temperature, and C is the determined correction factor, i.e. $\sim 0.1\%$. For a realistic temperature variation of $5°\,\text{K}$, the calculated correction is $\sim 150\,\text{eV}$ at $30\,\text{keV}$, a non-negligible value. It is not yet determined how often the temperature will be measured and if the temperature correction of the gain will be performed with every temperature measurement.

## 2.5 Energy addition and threshold comparison

To reconstruct the total energy of an event, all triggered anodes have to be taken into account. An event that is measured with two adjacent anodes can only be combined after the anodes have been converted to actual energy values, to reduce the discrepancies in the A/D conversion between the anodes. Only the total energy of the event is stored in the scientific event packets.

One of the final steps in the data pipeline is the confirmation that the event is indeed valid by comparing it with an upper and lower energy threshold. If the total energy of an event is above a predefined upper threshold, this event is treated as an high energy event and is discarded. Such events should be very rare, as in general they are early discarded due to the triggering of multiple anodes. If the total energy falls below the lower threshold value, the event is deemed to be not a significant detection and is also discarded. The rate of discarded events is counted and also included in the housekeeping data.

## 2.6 Energy scaling

The final step of the data processing in the MBEE is the energy reconstruction into a final 9 bit energy word. This reconstruction will be performed with a pre-determined, non-linear function with two energy regimes of:
$2\,\text{keV} - 30\,\text{keV}$: $60\,\text{eV}$ per Least-Significant Bit (LSB)
$30\,\text{keV} - 80\,\text{keV}$: $2\,\text{keV}$ per LSB

The final event package will consist of a 24 bit event with a 12 bit time stamp with a time resolution of $10\,\mu\text{s}$, a 9 bit energy word and a 3 bit event flag, indicating if the processed event was a single or double anode event. If the time between events is longer than the range allowed by the maximum 12 bit time stamp, dummy events are included to keep track of the absolute time.

A small data buffer is included in the MBEE to avoid a backlog of data from different processing pipelines before they are transmitted to the PBEE.

## 2.7 Housekeeping data

In addition to the scientific data, the MBEE is also responsible to transmit the housekeeping data (HK) to the PBEE and satellite bus. These data will be created in regular time intervals and can be used to monitor the health and performance of individual detectors. As part of the HK, the rates of valid and invalid events, as well as the different types of invalid events are monitored in order to correctly assess the dead time. The voltages for each SDD and the temperatures, which are needed for the gain correction, are also included in the HK.

## 2.8 Engineering mode

The above introduced processing pipeline describes the standard science mode and is necessary to obtain scientific data for each event. Due to the restriction in the telemetry to the ground, information of the detector and the anodes can not be included for each event. As there is no spatial resolution of the detector, such information is generally not relevant for the scientific goals. If more detailed information is required, either for onboard calibration purposes or to determine and narrow down a specific problem in one of the detectors, each individual module can be operated in an engineering mode. In this mode, the previously described data pipeline will be bypassed and information, such as the raw ADC values from all anodes, the trigger map, and the detector and ASIC information is transferred directly to the PBEE and the DHU. This data can then be used on the ground to update the gain calibration and the pedestal and threshold values in the MBEE or to turn off malfunction detectors. As this specific data mode creates a huge amount of data, only very short observations can be performed without overflowing the onboard memory.

The MBEE prototype is based on the ACTEL/RTAX 2000 FPGA. The IAAT has vast experience with this kind of FPGA, based on the heritage knowledge of the event-pre-processors for *XMM-Newton, e-Rosita, and IXO*.

## 3. PANEL BACK-END ELECTRONICS

The whole LAD instrument is divided into six individual panels with 21 modules, each. A Panel Back-End Electronics (PBEE) is mounted on each panel, handling all events from all 21 modules on this panel. The modules are independently connected to the PBEE so that a malfunctioning module would not influence the performance of the adjacent detectors. There is no redundancy at the PBEE level. A redundant PBEE would require to double all connections and cables on the panel, something which cannot be done due to mass and space restrictions. Although it is a critical point of failure for a whole detector panel, the loss of one of the six panels would not compromise the scientific goals of this mission.

The main tasks of the PBEE are:

- collecting and buffering the event packets from the 21 MBEEs
- introducing a differential timestamp to reduce the number of dummy events
- reformatting the data to binned data (see below) depending on the observation mode
- collection of HK data and creation of HK packets

### 3.1 Collecting the data

As the data pipeline in all MBEEs works synchronized with the onboard clock, only a small data buffer is included in the individual modules. To avoid conflicts or data losses during the transmission between the MBEEs and the PBEE, e.g. transmission of two simultaneous events, the data stream from all 21 MBEEs is collected and buffered individually before they are combined to one data stream and transmitted to the DHU for the final data compression.

### 3.2 Reformatting the timestamp

As each MBEE is using an individual differential time stamp, it is very important to keep track of the event time for each MBEE before they are combined to a single data stream. For weak sources, a lot of dummy time events are included in the data stream to keep track of the overall time. As these dummy time events do not contain any significant information and are only used to keep track of the time, the data from all 21 MBEEs can be combined to one data stream with a new, common differential time. In this way, the number of necessary dummy events decreases significantly, reducing also the amount of data which has to be transmitted to the ground.

As the limitation of bandwidth is primarily between the satellite and the ground station, we are exploring the possibility to use an absolute time for the individual MBEEs without the limitation of the 12 bit time. Such an absolute time would allow for more flexibility and significantly reduce the complexity in the reorganization of the individual events.

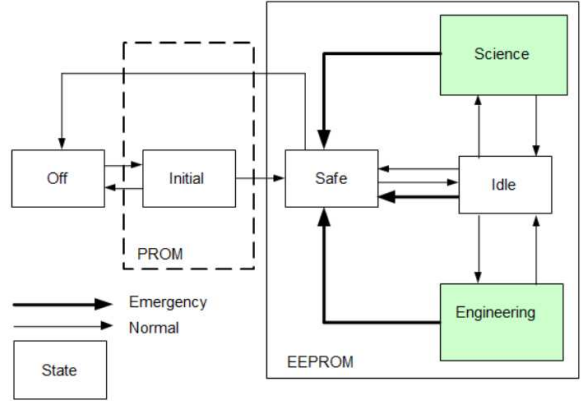

Figure 6. Different operational modes of a detector module, as defined by the DHU.

### 3.3 Creation of binned dataproducts

For an assumed average intensity of 500 mCrab, the estimated detector countrate in one module is $\sim$ 950 cts/s from one MBEE, which corresponds to 23 kbps of data. For the whole LAD with all 126 detector modules, the amount of data for bright sources (>500 mCrab) would exceed the available telemetry (assuming S-band telemetry). For this reason a pre-determined luminosity threshold is included in the DHU, which will switch the PBEE to an observer defined binned data mode, where the event-by-event data will be summarized in time resolved spectra and then discarded. The energy and time resolution of this spectrum is pre-defined by the observer and is only limited by the amount of total available bandwith, i.e. for a higher energy resolution the time resolution is reduced. In addition, the PBEE always creates a spectrum on a fixed defined energy and time resolution. This spectrum will be created for all sources and all luminosities, so that a comparison between individual observations can be achieved. The exact energy and timing resolution of these spectra, which will also be used for quicklook analysis, has not yet been determined.

### 3.4 Housekeeping

As mentioned above, the MBEE is responsible for measuring and transmitting the housekeeping packet for each detector to the PBEE. The PBEE accumulates the individual housekeeping packages from all modules and creates a housekeeping event, which will include valid and rejected event countrates, as well as the health information of each individual detector, i.e. voltages and temperatures. The PBEE creates a common housekeeping package for the whole panel and transmits the data to the DHU, where it is monitored.

The PBEE is also based on the experience gained in the *Athena*/HTRS instrument. We plan to use a Virtex-4 FPGA to implement the PBEE tasks.

## 4. DATA HANDLING UNIT

The data handling electronics (or data handling unit, DHU) forms the major controlling element of the LAD instrument. It provides the interface to the spacecraft and allows the control of the LOFT instrument subsystems. The DHU is built on a processor instead of an FPGA, chosen for its additional flexibility - when compared to a hardware only architecture, i.e. a state-machine.

The main functions of the DHU are:

- interfacing the PBEEs and the spacecraft onboard data handling unit
- instrument configuration
- data processing and compression

- HK health monitoring and calibration

## 4.1 Operational states

Figure 6 shows a summary of the different states of the DHU and in which order they are entered. Below are the different operational states described in more detail. The DHU can put every module independently into one of the different operational modes.

### 4.1.1 Booting

When the system is turned on for the first time, or rebooted, the DHU runs through an initial stage, where a bootstrap code is loaded from an onboard PROM. This mode is used to load, dump and run the main application code. The main application code is stored in an EEPROM on the satellite, but also can be loaded from the ground. Secondary rails and high voltages are off.

### 4.1.2 Safe

The safe mode is the first state entered when invoking the operational code. Initially only the DHU is on. All command and telemetry types are supported in the application code and the instrument can be safely powered off.

### 4.1.3 Idle

The idle software state is used whenever the instrument is not performing science or engineering tasks. It is important to note that even if the instrument is not acquiring data, the DHU remains turned on in order to further process already acquired data and to interact with the onboard data handling unit.

### 4.1.4 Science

In the science mode, the individual module operate as described above. The whole processing pipeline is executed and each event is either stored as individual events, or as counts in an integrated spectrum.

### 4.1.5 Engineering

The engineering operational mode is used for support operations on the detector modules. In this operational mode the measured data will not be processed through the MBEE data pipeline and will be directly compressed for transmission to ground. In addition, it is possible to perform a limited calibration (e.g.calculating a new offset/pedestal map) on the individual modules.

## 4.2 Health monitoring

All housekeeping data from the different panels are accumulated at the PBEE and then transmitted to the DHU. The DHU monitors the HK data and can switch individual modules into a safe mode, e.g. when then drift voltage collapses or if the detected temperature is outside of the operational range. The HK data are summarized and compressed before they are transmitted to the ground.

## 4.3 Data compression

As the LAD instrument creates a huge amount of data, a data compression system is unavoidable. Considering that we need to perfectly restore the measurements on ground, we must opt for a lossless approach. Based on the amount of necessary compression, a number of events will be bundled into an event package and then compressed before being sent to the ground. Finding a balance between a minimal required compression rate and the final size of the individual data packages is required, as a partial loss of the data package during transmission to ground results in a non-recoverable loss of the data.

We are working on a software data compression based on the High Time Resolution Spectrometer (HTRS) of the former International X-ray Observatory (IXO). The already developed electronics can easily be adapted to the detectors used in the LAD. A more detailed example of lossless data compression can be found in Wende et al. (2006).[9]

We are also investigating the possible application of the Prediction Error Coder (PEC) or Fully Adaptable PEC (FAPEC) entropy coders[7,8] on the energy and time data fields. The exact solution will depend on the computing resources available. These coders have proven to outperform the Rice coder when large fractions of outliers are present in the data. In the case of the energy information, we expect to have a rather steep statistic but with significant fractions of outliers — that is, several events with rather high energy values. On the other hand, the time information will probably have a rather flat statistic (depending on the events rate), for which PEC and FAPEC can also provide reasonable results very close to the Shannon limit. Preliminary tests confirm that ratios of 1.9 can be achieved on simulated energy data, whereas the Rice compressor reaches 1.8. The results for time data, on the other hand, are tightly coupled to the event rate. For 500 mCrab we just reach a negligible ratio of about 1.03 for any compressor (while the Shannon limit is about 1.1), whereas rates of 5 Crabs raises the ratios to almost 1.5. An adequate pre-processing stage is being designed in order to boost the achievable ratios. Also, using an adequate strategy of dummy events should decrease significantly the overall data volume. More frequent dummy events (which can be just three-bit markers) would reduce the number of bits required for the time data coding.

We are working on a software data compression based on the High Time Resolution Spectrometer (HTRS) of the former International X-ray Observatory (IXO). The already developed electronics can easily be adapted to the detectors used in the LAD. A more detailed example of lossless data compression can be found in Wende et al. (2006).[9]

## 5. SUMMARY

Although the general consent of the data pipeline in the LOFT LAD instrument is well defined, a number of different trade-offs and design possibilities are not yet fixed. The vast number of necessary ASICs for the readout requires a very simple and modular data processing scheme, which can be easily changed. As the ASIC will be designed from scratch, necessary features can be easily included, if they do not increase too much complexity and power consumption.

The Wide Field Monitor (WFM), the second scientific instrument on LOFT, is using the same detectors to readout.[3] The main difference is that the WFM is a coded-mask instrument, using a much finer anode pitch to allow a spatial resolution in one dimension. The MBEE processing pipeline can easily be adapted to read out the WFM detectors, with the main difference that more anodes will trigger at the same time.

The work at University Tübingen includes the development of a full prototype for the MBEE data-pipeline process by the end of this year. This processing pipeline will be fully modular, so that it can easily be adapted to changes in the existing data format. Experience in an event-pre-processing pipeline has been accumulated through a large number of different missions, i.e. *e-Rosita*, *XMM-Newton*, and *IXO* The PBEE and the compression in the DHU are based on heritage from previous missions, i.e. *IXO*, and can be easily adapted to the LAD data stream. The IEEC has a prototype implementation of PEC and FAPEC in an ACTEL ProASIC FPGA, as well as prototype software of PEC that has been successfully tested on the data processors that will fly onboard the Gaia mission.[10]

## Acknowledgements

This work is partially supported by the *Bundesministerium für Wirtschaft und Technologie* through the *Deutsches Zentrum für Luft- und Raumfahrt e.V. (DLR)* under the grant number FKZ 50 OO 1110. The Italian teams are supported by ASI, INAF and INFN. The work at MSSL is supported by the UK Space Agency. The work at the IEEC is supported by the MICINN-FEDER through grant AYA2009-14648-C02-01 and CONSOLIDER CSD2007-00050.